\documentclass[aps,prl,groupedaddress,showpacs]{revtex4}

\usepackage{amsfonts}
\usepackage{graphicx}

\begin{document}


\title{Exact Massive Solutions of Classical Massless Field Theories}


\author{Marco Frasca}
\email[]{marcofrasca@mclink.it}
\affiliation{Via Erasmo Gattamelata, 3 \\ 00176 Roma (Italy)}


\date{\today}

\begin{abstract}
We give the exact solution of classical equation of motion of a quartic scalar massless
field theory showing that this is massive and is represented by a superposition of free
particle solutions with a discrete spectrum. Then we show that this is also a solution of
the classical Yang-Mills field theory that is so proved acquiring mass by dynamical evolution
with a corresponding discrete mass spectrum. Finally we develop quantum field theory starting with this solution.
\end{abstract}


\maketitle


There are two main reasons why getting an exact solution to a classical field theory is an important step beyond. A classical solution may give a clever understanding of the physics underlying the theory itself and secondly, but not less important, one can build a quantum field theory out of such a solution using perturbation theory. Indeed, somewhat striking may happen already at a classical level for a classical field theory. A very blatant example of this is given by spontaneous breaking of symmetry. We see, already at the classical level, a non trivial behavior of the theory that has as an important side effect mass generation for all other coupled fields.

Generally speaking, we do not have many of such exact solutions. What we do most of times is perturbation theory starting from the solutions of the free theory that, being linear, is straightforwardly solved. For a lot of interesting cases our ability is just limited by the fact that we have to cope with non-linear equations whose solutions are rarely known.

Recently, we published a paper where we were able the obtain both the spectrum and the Feynman propagator in the infrared of a scalar theory \cite{fra1}. Our aim here is to reconsider this theory in the classical limit to obtain the unexpected results that the behavior of the exact classical solution is the same of the quantum field theory in the infrared we derived giving in this way a consistent mathematical support to our previous conclusions.

So, we start by writing down the action for the theory we aim to solve. One has
\begin{equation}
    S = \int d^4x\left[\frac{1}{2}(\partial\phi)^2-\frac{\lambda}{4}\phi^4\right].
\end{equation}
We just note that for this theory holds conformal invariance having no mass term and does not have any dimensional parameter being $\lambda$, the coupling, adimensional. So, Euler-Lagrange equation reduces in this case to the equation of motion
\begin{equation}
\label{eq:phi}
    \ddot\phi-\Delta_2\phi+\lambda\phi^3=0.
\end{equation}
We can write down immediately an exact solution of this equation being given by
\begin{equation}
\label{eq:phis}
    \phi(x)=\mu\left(\frac{2}{\lambda}\right)^{\frac{1}{4}}{\rm sn}(p\cdot x+\theta,i)
\end{equation}
being ${\rm sn}$ the Jacobi snoidal function, $\mu$ and $\theta$ two integration constants. The first has the dimension of a mass while the other, being just a phase, is adimensional. The important point is that this solution holds only if the following dispersion relation holds
\begin{equation}
\label{eq:disp}
    p^2=\mu^2\left(\frac{\lambda}{2}\right)^{\frac{1}{2}}
\end{equation}
that is we have a wave-like solution with a mass $m_0=\mu\left(\frac{\lambda}{2}\right)^{\frac{1}{4}}$ that we see going to zero with the coupling going to zero. So, we have reached a striking conclusion that, starting with a massless theory, the corresponding classical solution is indeed massive.

Let us now get into the solution (\ref{eq:phis}) to understand what is going on. This solution represents a nonlinear wave solution that is generally well-known in physics. But in order to understand the physics of the field theory we use the following Fourier expansion, a 
standard result of Jacobi elliptic functions \cite{gr},
\begin{equation}
    {\rm sn}(u,i)=\frac{2\pi}{K(i)}
    \sum_{n=0}^\infty\frac{(-1)^ne^{-(n+\frac{1}{2})\pi}}{1+e^{-(2n+1)\pi}}
    \sin\left[(2n+1)\frac{\pi u}{2K(i)}\right]
\end{equation}
being $K(i)$ the following elliptic integral
\begin{equation}
    K(i)=\int_0^{\frac{\pi}{2}}\frac{d\theta}{\sqrt{1+\sin^2\theta}}\approx 1.3111028777.
\end{equation}
This means that the exact solution is given by a superposition of plane waves as
\begin{equation}
    \phi(x)=-\mu\left(\frac{2}{\lambda}\right)^{\frac{1}{4}}\frac{\pi}{iK(i)}
    \sum_{n=0}^\infty\frac{(-1)^ne^{-(n+\frac{1}{2})\pi}}{1+e^{-(2n+1)\pi}}
    \exp\left[-i(2n+1)\frac{\pi p\cdot x}{2K(i)}\right]+c.c.
\end{equation}
that is, the field is just a set of excitations typical of free particles having a discrete set of eigenvalues. Indeed, we can single out a mass spectrum. This can be easily accomplished in the rest frame setting ${\bf p}=0$. We are left with the following field
\begin{equation}
    \phi(t,0)=-\mu\left(\frac{2}{\lambda}\right)^{\frac{1}{4}}\frac{\pi}{iK(i)}
    \sum_{n=0}^\infty\frac{(-1)^ne^{-(n+\frac{1}{2})\pi}}{1+e^{-(2n+1)\pi}}
    \exp\left[-i(2n+1)\frac{\pi}{2K(i)}
    \left(\frac{\lambda}{2}\right)^{\frac{1}{4}}\mu t\right]+c.c.
\end{equation}
where use has been made of the dispersion relation (\ref{eq:disp}). So we have got the mass spectrum
\begin{equation}
    \epsilon_n=(2n+1)\frac{\pi}{2K(i)}
    \left(\frac{\lambda}{2}\right)^{\frac{1}{4}}\mu
\end{equation}
that coincides exactly with the one obtained through the quantum field theory presented in \cite{fra1}. We see that already at the classical level the theory appears trivial. We have just a superposition of plane waves with the discrete spectrum of a harmonic oscillator, notwithstanding we started with a non-trivial non-linear theory. But what is really striking is the appearing of a mass spectrum out of a classical massless field theory. This means that conformal invariance is dynamically broken. Finally we note that a relatively strong coupling is needed to break conformal invariance. The reason for this relies on the fact that the mass that appears in our solution goes to zero with the fourth root of the coupling. We note that we can identify a set of ``golden numbers'' through the ratio $\epsilon_n/m_0$. The first ones are given in table \ref{tab:gn}.
\begin{table}
\begin{tabular}{|c|c|c|c|} \hline\hline
Order & Value\\ \hline
0 & 1.198140235 \\ \hline 
1 & 3.594420705 \\ \hline
2 & 5.990701175 \\ \hline
3 & 8.386981645 \\ \hline\hline
\end{tabular}
\caption{\label{tab:gn} The first set of ``golden numbers''.}
\end{table}

On the same ground we can consider a scalar theory with a broken phase. We can write down
for this case as
\begin{equation}
    S = \int d^4x\left[\frac{1}{2}(\partial\phi)^2+
    \frac{1}{2}m^2\phi^2-\frac{\lambda}{4}\phi^4\right]
\end{equation}
giving the equation
\begin{equation}
\label{eq:phib}
    \ddot\phi-\Delta_2\phi+ \lambda\phi^3-m^2\phi=0.
\end{equation}
It is straightforward to write down the exact solution as
\begin{equation}
    \phi(x)=v\cdot{\rm dn}(p\cdot x+\theta,i)
\end{equation}
being ${\rm sn}$ the Jacobi snoidal function,
$v=\sqrt{2m^2/3\lambda}$ and $\theta$ an integration constant. Also in this case, such a
solution holds only when the following dispersion relation holds
\begin{equation}
     p^2=\frac{\lambda v^2}{2}.
\end{equation}
In this case we have a zero mass excitation into the spectrum. Indeed, if we write the Fourier expansion, that is a well-known result of Jacobi elliptic functions \cite{gr}, as
\begin{equation}
     {\rm dn}(u,i)=\frac{\pi}{2K(i)}+\frac{2\pi}{K(i)}\sum_{n=1}^\infty
     \frac{e^{-n\pi}}{1+e^{-2n\pi}}\cos\left(2n \frac{\pi u}{2K(i)}\right),
\end{equation}
one has
\begin{equation}
     \phi(x)=\frac{v\pi}{2K(i)}+\frac{v\pi}{K(i)}\sum_{n=1}^\infty
     \frac{e^{-n\pi}}{1+e^{-2n\pi}}e^{-2in \frac{\pi}{2K(i)}px}+c.c.
\end{equation}
and, as above, setting ${\bf p}=0$ we get
\begin{equation}
     \phi(t,0)=\frac{v\pi}{2K(i)}+\frac{v\pi}{K(i)}\sum_{n=1}^\infty
     \frac{e^{-n\pi}}{1+e^{-2n\pi}}e^{-2in 
     \sqrt{\frac{\lambda}{2}}\frac{\pi v}{2K(i)}t}+c.c.
\end{equation}
and this gives the spectrum
\begin{equation}
     \epsilon_n=n \frac{\pi}{K(i)}\sqrt{\frac{\lambda}{2}}v
\end{equation}
with $n=0,1,2,\ldots$. These results show how a classical scalar theory with a wrong mass sign
recovers a proper physical mass spectrum due to the non-linear term. This result confirms our derivation given in Ref.\cite{fra2}

In order to have an idea of the physical meaning of these classical solutions, we point out that are the same of plane waves of a free theory and so boundary conditions are those generally considered in this case. This can be seen when the Fourier series of these solutions is obtained showing a super-position of plane waves. So, identical boundary conditions should apply. Changing these conditions, we come to different physical situations and so to a different spectrum of excitations.

We want to see, in the same way as done for the scalar theory, if the classical Yang-Mills theory admits a massive solution. We proceed like for the scalar field and write down the action that now is given by
\cite{nair}
\begin{equation}
    S=\int d^4x\left[\frac{1}{2}\partial_\mu A^a_\nu\partial^\mu A^{a\nu}
    +gf^{abc}\partial_\mu A_\nu^aA^{b\mu}A^{c\nu}
	+\frac{g^2}{4}f^{abc}f^{ars}A^b_\mu A^c_\nu A^{r\mu}A^{s\nu}\right].
\end{equation}
We can map this action on the one of the massless quartic scalar field so that a
solution of the equation of motion of the scalar field is also a solution of 
the Yang-Mills equations of motion \cite{fra3,fra4}. E.g., for SU(3), the following choice $A_\mu^a=((0,0,0,0),(0,\phi,0,0),(0,\phi,0,0),(0,0,\phi,0),(0,\phi,\phi,0),(0,0,\phi,0),(0,0,0,\phi),(0,0,0,\phi))$ is a possible mapping but the number of choices (Smilga's choices) is truly large and dependent on the gauge group.
Given these choices, Yang-Mills action reduces to the one of the scalar field (\ref{eq:phi}) with the substitution $\lambda\rightarrow Ng^2=3g^2$ that is the 't~Hooft coupling. We get, as it should be, $N^2-1=8$ of such equations. So, by introducing the tensor $\eta^a_\mu=((0,0,0,0),(0,1,0,0),(0,1,0,0),(0,0,1,0),(0,1,1,0),(0,0,1,0),(0,0,0,1),(0,0,0,1))$, we can write down our solution as
\begin{equation}
    A^a_\mu(t,0)=\eta^a_\mu\phi(t)=\eta^a_\mu\Lambda
    \left(\frac{2}{3g^2}\right)^{\frac{1}{4}}
    {\rm sn}\left(\left(\frac{3g^2}{2}\right)^{1 \over 4}\Lambda t+\varphi,i\rm\right)
\end{equation}
being now $\Lambda$ and $\varphi$ the arbitrary integration constants. After a Lorentz transformation $\Lambda^\nu_\mu$ we will get
\begin{equation}
    A^a_\mu(x)=\Lambda^\nu_\mu\eta^a_\nu\phi(x)=\hat\eta^a_\mu\Lambda
    \left(\frac{2}{3g^2}\right)^{\frac{1}{4}}{\rm sn}(p\cdot x+\varphi,i\rm)
\end{equation}
and the dispersion relation
\begin{equation}
\label{eq:disp2}
    p^2=\Lambda^2\left(\frac{3g^2}{2}\right)^{\frac{1}{2}}
\end{equation}
that proves our assertion that the classical Yang-Mills equations admit massive solutions. Being all exactly as for the scalar field, we get again a mass spectrum of free particles given by
\begin{equation}
    \epsilon_n^{YM}=(2n+1)\frac{\pi}{2K(i)}
    \left(\frac{3g^2}{2}\right)^{\frac{1}{4}}\Lambda.
\end{equation}
So, by dividing for the mass as given in (\ref{eq:disp2}), we get the same set of ``golden numbers'' as those of the scalar field (see tab. \ref{tab:gn}). This mass of the Yang-Mills field is usually identified with the square root of the string tension $\sigma$ and its value is either $440$ MeV or $410$ MeV depending on the analysis carried out by different groups (e.g. \cite{tep,mor}). 
The spectrum we obtain differs from what quantum expectations could be. Indeed, one cannot see asymptotic freedom being this originating from quantum corrections. As such these could modify the spectrum that eventually could be maintained just in the infrared, an evidence to be shown. One can recover the spectrum of a free particle at very high energies just working with the Fourier transform of the given exact solution.

In order to work out a quantum field theory starting from the above exact solutions, we consider as usual the generating functional
\begin{equation}
    Z[j] = N\int[d\phi] e^{i\int d^4x
    \left[\frac{1}{2}(\partial\phi)^2-\frac{\lambda}{4}\phi^4+j\phi\right]}
\end{equation}
being $N$ a normalization constant, and we take the substitution $\phi=\phi_c+\delta\phi+O(\delta\phi^2)$ being $\phi_c$ the classical solution given in eq.(\ref{eq:phis}). We will recover the results given in \cite{fra1} and the first higher correction as it should be. After the substitution has done one has
\begin{equation}
    Z[j]=e^{i[S_c+\int d^4x j\phi_c]} \int[d\delta\phi] e^{i\int d^4x
    \left[{1 \over 2}(\delta\phi)^2
    -{3\over 2}\phi_c^2(\delta\phi)^2+j\delta\phi\right]}+O((\delta\phi)^3).
\end{equation}
Before to evaluate the higher order correcion, we take a look at the term
\begin{equation}
    Z_0[j]=e^{i[S_c+\int d^4x j\phi_c]}.
\end{equation}
We are able to put this term into an useful form if we are able to solve the equation
\begin{equation}
    \partial^2_x\Delta_0(x)+\lambda \Delta_0^3(x)=\delta^4(x)
\end{equation}
and this can be easily done with eq.(\ref{eq:phis}) and introducing Heaviside function. But for our needs, the limit $\lambda\rightarrow\infty$, it is enough to consider a small time expansion writing down the solution as \cite{fra5,fra6}
\begin{equation}
    \phi_c(x)\approx\int d^4y\Delta_0(x-y)j(y)
\end{equation}
yielding
\begin{equation}
    Z_0[j]\approx e^{\frac{i}{2}\int d^4xd^4yj(x)\Delta_0(x-y)j(y)},
\end{equation}
in agreement with Ref.\cite{fra1}, reducing the theory to the Gaussian case. Finally, in order to compute the next to leading order correction, we do the following change of coordinates
\begin{equation}
    \delta\phi=\delta\phi_0+\int d^4y\Delta_1(x-y)j(y),
\end{equation}
being
\begin{equation}
\label{eq:d1}
    \partial^2_x\Delta_1(x-y)+3\lambda\phi_c^2(x)\Delta_1(x-y)=\delta^4(x-y)
\end{equation}
the first order propagator, and we are left with a Gaussian functional
\begin{equation}
    Z[j]\approx e^{{i\over 2}\int d^4xd^4yj(x)[\Delta_0(x-y)+\Delta_1(x-y)]j(y)}.
\end{equation}
In order to compute the first order propagator, we have to solve eq.(\ref{eq:d1}). This can be done noting that, in the limit $\lambda\rightarrow\infty$ one can use a WKB approximation in a gradient expansion. We get the equation
\begin{equation}
\label{eq:d2}
    \partial^2_t\Delta_1(x-y)+3\lambda\phi_c^2(x)\Delta_1(x-y)\approx\delta(t-t_y)\delta^3(x-y)
\end{equation}
and the approximate solution
\begin{equation}
    \Delta_1(x-y)=\theta(t-t_y)\delta^3(x-y)\frac{1}{(3\lambda)^{1\over 4}
    \phi_c^{1\over 2}(x)}e^{-i\sqrt{3\lambda}
    \int_{t_y}^tdt'\phi_c({\mathbf x},t')}+
    \theta(t_y-t)\delta^3(x-y)\frac{1}{(3\lambda)^{1\over 4}
    \phi_c^{1\over 2}(x)}e^{i\sqrt{3\lambda}
    \int_{t_y}^tdt'\phi_c({\mathbf x},t')}.
\end{equation}
This solution is just an approximation that should be improved with higher order corrections to recover full Lorentz invariance. It is interesting to note that this correction goes like $1/\lambda^{1\over 4}$ showing that we obtained a strong coupling expansion that holds in the limit $\lambda\rightarrow\infty$. Some special treatment is required for the poles in this Green function due to the solutions of $\phi_c(x)=0$. These are known as caustics.

We have shown in this paper how a self-interacting massless field can generate massive excitations and we have obtained this through exact solutions of classical equations of motion. A mass pole is originating by the non-linearity of the theory and it is just a dynamical effect. This classical solution is recovered, as we have already shown \cite{fra1}, in quantum field theory working the other way around.
 


\begin{thebibliography}{99}
\bibitem{fra1} M. Frasca, Phys. Rev. D {\bf 73}, 027701 (2006); Erratum-ibid. D {\bf 73}, 049902 (2006).
\bibitem{gr} I. S. Gradshteyn, I. M. Ryzhik, {\sl Table of Integrals, Series, and Products},
(Academic Press, 2000).
\bibitem{fra2} M. Frasca, Int. J. Mod. Phys. A {\bf 22}, 5345 (2007).
\bibitem{nair} V. P. Nair, {\sl Quantum Field Theory}, (Springer, New York, 2005), p.194.
\bibitem{fra3} M. Frasca, Phys. Lett. {\bf B670}, 73 (2008).
\bibitem{fra4} M. Frasca, arXiv:0903.2357 [math-ph].
\bibitem{tep} B. Lucini, M. Teper, U. Wenger, JHEP {\bf 06}, 012 (2004).
\bibitem{mor} Y. Chen, A. Alexandru, S. J. Dong, T. Draper, I. Horvath, F. X. Lee, K. F. Liu, N. Mathur, C. Morningstar, M. Peardon, S. Tamhankar, B. L. Young, J. B. Zhang, Phys. Rev. D {\bf 73}, 014516 (2006).
\bibitem{fra5} M. Frasca, Mod. Phys. Lett. A {\bf 22}, 1293 (2007).
\bibitem{fra6} M. Frasca, Int. J. Mod. Phys. A {\bf 23}, 299 (2008). 
\end{thebibliography}
\end{document}